\documentclass[pra,twocolumn,superscriptaddress,showpacs]{revtex4} 
\usepackage[english]{babel}
\usepackage[dvips]{graphicx}
\usepackage{amssymb}
\usepackage{amsmath}
\usepackage{color}
\begin{document}

\title{Lasing and transport in a quantum dot-resonator circuit}

\author{Pei-Qing Jin}
\affiliation {Institut f\"ur Theoretische Festk\"orperphysik,
      Karlsruhe Institute of Technology, 76128 Karlsruhe, Germany}
\author{Michael Marthaler}
\affiliation {Institut f\"ur Theoretische Festk\"orperphysik,
      Karlsruhe Institute of Technology, 76128 Karlsruhe, Germany}
\author{Jared H.~Cole}
\affiliation {Institut f\"ur Theoretische Festk\"orperphysik,
      Karlsruhe Institute of Technology, 76128 Karlsruhe, Germany}
\affiliation {Applied Physics, School of Applied Sciences,
RMIT University, Melbourne 3001, Australia}
\author{Alexander Shnirman}
\affiliation{Institut f\"{u}r Theorie der Kondensierten Materie,
Karlsruhe Institute of Technology, 76128 Karlsruhe, Germany}
\affiliation {DFG Center for Functional Nanostructures (CFN),
      Karlsruhe Institute of Technology, 76128 Karlsruhe, Germany}
\author{Gerd Sch\"on}
\affiliation {Institut f\"ur Theoretische Festk\"orperphysik,
      Karlsruhe Institute of Technology, 76128 Karlsruhe, Germany}
\affiliation {DFG Center for Functional Nanostructures (CFN),
      Karlsruhe Institute of Technology, 76128 Karlsruhe, Germany}

\date{\today}

\begin{abstract}
We study a double quantum dot system coherently coupled to an
electromagnetic resonator. A current through the dot system can
create a population inversion in the dot levels and, within a narrow resonance window,
 a lasing state in the resonator. The lasing state correlates with the transport properties.
On one hand, this allows probing the lasing state via a current measurement.
On the other hand, the resulting narrow current peak allows
resolving small differences in the dot properties, e.g., a small difference in the
Zeeman splittings of the two dots. For realistic situations relaxation processes
have pronounced consequences. Remarkably, they may even enhance the resolution
between different spin states by releasing a trapped population in the off-resonant spin channel.
\end{abstract}

\pacs{42.50.Pq, 73.21.La, 03.67.Lx}

\maketitle

\section{Introduction}

 Atomic levels coupled to electromagnetic fields lead to a variety of
 fundamentally important quantum effects.
 A prime example is the laser, created by exciting atoms in a cavity which are coupled to
 a radiation field. Many further basic effects have been studied over the years
 in the field of quantum optics and quantum electrodynamics (QED).
 Recently, similar effects were demonstrated in solid states systems.
 In these ``circuit QED"-setups superconducting qubits, serving as artificial
 two-level systems, are coupled to superconducting electromagnetic circuits
 \cite{Wallraff04,Chiorescu04,Blais04,Schoelkopf}. On one hand, many of the effects predicted for
  quantum optics systems could be demonstrated with unprecedented quality. On the other hand,
 the new parameter regime, i.e., strong coupling, low temperature, and single-qubit rather
 than many atoms, revealed also
 qualitatively novel behavior. An example is lasing with a single qubit
 observed recently \cite{Astafiev,hauss08,grajcar}, where
 for strong coupling to the oscillator, quantum noise influences
 the linewidth of the emission spectrum in a characteristic way \cite{SQL1,SQL2,SQL3,Wallraff2010}.

In this paper we propose a different circuit QED setup, where the
superconducting qubit is replaced by a semiconductor double quantum
dot with discrete energy levels. It is coupled to an electromagnetic
resonator, preferably a superconducting one in order to achieve a
high quality factor. Double quantum dots
with different charge states can serve as realizations
of quantum bits, and indeed both single-qubit coherent manipulations
and two-qubit operations have been demonstrated
\cite{Hayashi,Petta04,Gorman,Shinkai}.
Moreover, quantum-dot lasers that operate in the optical regime
were realized and, e.g., anti-bunching of photons has been observed \cite{Wiersig,Su}.
By placing the dot system
between two electrodes and driving a current through the system it
is straightforward to create a population inversion, which then can
lead to a lasing state in the resonator with frequency in the few GHz range \cite{Childress}.

The lasing state in the resonator correlates with the transport properties
through the double dot system.
On one hand, this allows probing the lasing state via a current measurement,
which may be easier to perform in an experiment.
On the other hand, the resulting narrow current peak allows
resolving small differences in the dot properties.
This opens perspective for applications of the setup for high resolution measurements.
As an example we will analyze the consequences of a small difference in the
Zeeman splittings of the two dots.

The shorter relaxation and coherence times of charge degrees of freedom in
semiconductor quantum dots, as compared to spin-based or superconducting qubits,
makes them less interesting for applications of quantum
information processing. However, for lasing and other
signatures of the coupling to the resonator to be studied here, the
requirements on the coherence time are less stringent. Still,
relaxation and decoherence processes have strong effects, and much
of the following is devoted to their analysis.

Spin degrees of freedom in semiconducting quantum dots promise long
coherence times, and indeed were the target of earlier work
\cite{Burkard06,Trif,Cottet}.
Particularly long-lived are logical qubits realized by singlet and
triplet states of doubly occupied double quantum dot systems.
For manipulations of these spin qubits,
the nuclear-spin induced Zeeman splitting difference, of the order of
several millitesla between the two dots, was exploited \cite{Petta05}.
In this paper we suggest to probe and resolve this small difference
with the help of the sharp lasing resonance condition. It leads to
two separate peaks in the photon number and the transport current as
a function of detuning. Surprisingly, the resolution of these resonances
can actually be enhanced by relaxation effects, since they can release
population trapped in the off-resonant spin channel which otherwise
blocks transport.

Clearly the setup proposed in this paper bears similarities to 
bandgap driven semiconductor lasers as well as quantum cascade lasers \cite{Hall,Faist}.
However, there are crucial differences in the parameter regime
where the systems are operated.
The present system operates in the GHz regime (hence we should actually talk about a ``maser")
as compared to the THz regime of conventional lasers.
The levels of the dots driven to show the population inversion are sharply defined
and can be tuned by gate voltages.
Also the coupling strength and tunneling rates can be varied {\sl in situ}
\cite{Tarucha,Oosterkamp,Fujisawa}.
The double-dot system has a large dipole moment leading to a strong 
coupling to the resonator, which is needed to create a ``single-atom maser".
By operating at low temperatures and by using a high-Q resonator 
we minimize dissipative effects.
The combination of these effects creates the sharp resonance condition 
which is the focus of this work.

 The paper is organized as follows. In Sec. \ref{Sec:Charge}, we present the model
 for the quantum dots coupled to the resonator and describe the pumping to create a population inversion.
 The theoretical tools for the analysis of relaxation and dephasing are
 also introduced. We analyze the photon population of the resonator and
 correlate it to transport and fluctuations properties of the dot system.
 In Sec. \ref{Sec:Spin}, we extend the model to include a spin splitting
 and find a rich
 dependence on various control parameters of the system, which can be resolved
 due to the narrow resonance condition for lasing.
 We conclude with a brief summary in Sec. \ref{Sec:Summary}.


\section{Charge states}\label{Sec:Charge}

\subsection{Model and method}

\begin{figure}[t]
 \centering
 \includegraphics[width=0.3\textwidth]{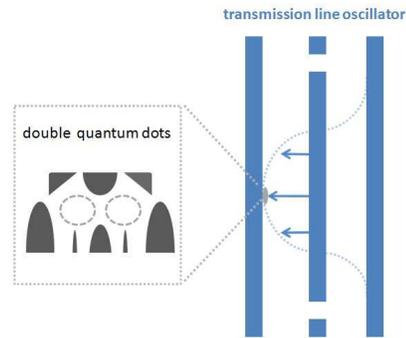}
 \caption{(Color online) Illustration of a double quantum dot-resonator
 circuit. The dot is placed at a maximum of the electric field of the transmission line
  in order to maximize the  dipole interaction with the resonator.
  For gate-defined quantum dots most parameters can be tuned by the applied voltages. }\label{fig:dot}
 \end{figure}

 We consider a semiconducting double quantum dot system
 with large charging energy and discrete energy levels
 coupled to a high-$Q$ electromagnetic resonator. The former can be created and controlled by lateral
 structuring of a two dimensional electron gas (2DEG) by applied gate voltages, the latter can be realized by a superconducting transmission line as shown schematically in Fig.~\ref{fig:dot}.
 The dot system is biased such that the two relevant basis states are
 $|1,0\rangle$ and $|0,1\rangle$ with a
 single electron occupying either the left or right dot, respectively.
 These states are referred to in the following as pure charge states.
 They are assumed to have a bare energy difference $\epsilon$, but additionally they
 are coupled by coherent interdot tunneling with strength $t$.
 In experiments, both parameters can be tuned in the range from several to tens of $\mu{\rm eV}$
 \cite{Oosterkamp,Harbusch}, which fits the parameter regime considered in this paper.
 Hence the Hamiltonian of the double dot system is
 \begin{eqnarray}\label{eq_Hamiltonian_in_Charge_Basis}
 H_{\rm dd} = \frac{1}{2} (\epsilon\, \tau_z +~ t\, \tau_x),
 \end{eqnarray}
 with Pauli matrices $\tau_x= |1,0\rangle\langle 0,1|+|0,1\rangle\langle 1,0|$
 and similar for $\tau_z$.
 Transport through the double dot system involves a third state; below we will
 assume it to be the  empty-dot state $|0,0\rangle$.

 The transmission line can be modeled as a harmonic oscillator with frequency $\omega_{\rm r}$,
 which in most circuit QED experiments is of the order of 1 to 10 GHz \cite{Wallraff04}.
 In the arrangement shown in Fig. \ref{fig:dot}, the two pure charge states
 have different dipole moments and couple to the resonator. The interaction is
 \begin{eqnarray}
 H_{\rm I} = \hbar g_0 (a^\dag+a^{}) \tau_z,
 \end{eqnarray}
 where $a^{}$ ($a^\dag$) represents the annihilation (creation) operator of photons in the resonator.
 For weak interdot tunneling, when the single-particle wavefunctions are strongly localized in either dot,
 the coupling strength can be estimated as $g_0 \sim eEd/(2\hbar)$
 with $d$ being the distance between the centers of the two dots.
 The electric field at the antinode of the resonator mode can achieve values of order
 $E \sim 0.2~\rm{V}/\rm{m}$,
 which for a distance $d \sim 0.3~\rm{\mu m}$ leads to a coupling strength $g_0\sim 50~{\rm MHz}$  \cite{Wallraff04}.

To proceed we introduce the eigenbasis of the Hamiltonian (\ref{eq_Hamiltonian_in_Charge_Basis}),
 \begin{eqnarray}\label{eq_Eigenbasis_of_the_Dot}
 |e\rangle &=& \cos\left(\theta/2\right)|1,0\rangle + \sin\left(\theta/2\right)|0,1\rangle,
               \nonumber \\
 |g\rangle &=& -\sin\left(\theta/2\right)|1,0\rangle + \cos\left(\theta/2\right)|0,1\rangle,
 \end{eqnarray}
 with the angle $\theta = \arctan(t/\epsilon)$ characterizing the mixture of the pure charge states.
 At the  point of degeneracy, $\epsilon=0$, the mixing angle is  $\theta = \pi/2$.
 In the basis (\ref{eq_Eigenbasis_of_the_Dot}), within the rotating wave approximation,
 the Hamiltonian for the coupled system reduces to
 \begin{eqnarray}\label{eq:ChargeH}
 H_{\rm{sys}}  = \frac{\hbar\omega_0}{2} \sigma_z
 + \hbar \omega^{}_{\rm r} a^\dag a
 + \hbar g (a^\dag \sigma_- +a^{} \sigma_+).
 \end{eqnarray}
 Here $\omega_0 = \sqrt{\epsilon^2+t^2}/\hbar$ denotes the frequency of the two-level system.
 It can be tuned via gate voltages, which allows for a detuning $\Delta = \omega_0-\omega_{\rm r}$
 from the resonator frequency.
 The coupling strength, $g = g^{}_0 \sin\theta$, reaches its maximum at the degeneracy point
 and decreases as one moves away from this point.

 We analyze the dynamics of the coupled dot-resonator system, which is assumed
 to be coupled  weakly to an environment with smooth spectral function, in the frame of
 a master equation for the reduced density matrix $\rho$
 in the  Born-Markovian approximation \cite{Gardiner,Carmichael}.
 Throughout this paper we consider low temperatures, $T =0$, with vanishing
 thermal photon number and excitation rates.
 In this case, the master equation is
 \begin{eqnarray} \label{eq:ME}
 \dot \rho &=& -\frac{i}{\hbar}\left[H_{\rm_{sys}}, \rho\right]
  +\mathcal L_{\rm r}\, \rho + \mathcal L_{\rm \downarrow}\rho+\mathcal L_{\rm \varphi}\, \rho
  + \mathcal L_{\rm L}\, \rho+\mathcal L_{\rm R}\, \rho
                            \nonumber \\
 &\equiv& \mathcal L_{\rm tot} \, \rho.
 \end{eqnarray}
 The dissipative dynamics is described by Lindblad operators of the form
 \begin{equation}
 {\cal L}_{i}\rho=\frac{\Gamma_i}{2}\left(2L_i\rho L_i^{\dag}-L_i^{\dag}L_i\rho-\rho L_i^{\dag}L_i\right).
 \end{equation}
 For the oscillator we take the standard decay terms $L_{\rm r}=a$ with rate $\Gamma_{\rm r}=\kappa$.
 For the two-level system we account for the relaxation by $L_\downarrow =\sigma_-$ with rate
 $\Gamma_\downarrow$
 and for decoherence by $L_{\varphi}=\sigma_z$ with rate $\Gamma_\varphi^*$.
 The last two terms $ \mathcal L_{\rm L/R}$ account for the incoherent tunneling between the
 electrodes and the left and right dots.

 To achieve a lasing state, a low decay rate of the resonator $\kappa$ is required \cite{SQL1,SQL2},
 satisfying
 \begin{eqnarray}\label{eq:lasing}
 \kappa < \frac{2g^2\tau_0}{\Gamma_\varphi^* + \Gamma_\downarrow/2},
 \end{eqnarray}
 with $\tau_0$ the population inversion to be introduced later.
 Hence the resonator should have a high Q-factor. The required quality can be reached with a
 superconducting transmission line, for which Q $\sim10^6$ has been demonstrated \cite{Barends}.
 We note that also the coupling between superconducting leads and semiconductor quantum dots
 has been demonstrated, e.g., in the Cooper pair splitter experiments \cite{Hofstetter}.
 Throughout this paper, we choose the Q-factor to be $10^5$, corresponding to a decay rate
 $\kappa = 10^{-5} \omega_{\rm r}$.

\begin{figure}[t]
 \centering
 \includegraphics[width=0.35\textwidth]{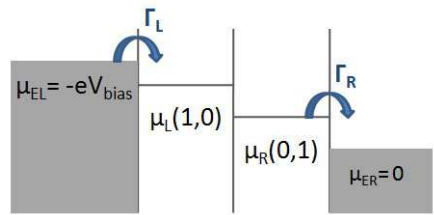}
 \caption{(Color online) Tunneling sequence in the double dot system.
 The chemical potentials $\mu_{\rm L} (1,0)$ and $\mu_{\rm R} (0,1)$ of the two dots
 are assumed to be arranged as indicated.}\label{fig:pump}
 \end{figure}

 A crucial prerequisite for lasing is the pumping described in (\ref{eq:ME})
 by the incoherent tunneling terms  $ \mathcal L_{\rm L/R}$.
 As in an optical laser the pumping involves a third state; for the bias which we consider
 this is the empty state $|0,0\rangle$ of the double dot.
 We further assume that the system is biased
 such that only the chemical potentials of the states $|1,0\rangle$ and $|0,1\rangle$
 lie within the window defined by the drain and source voltages, and they are arranged as
 shown in Fig. \ref{fig:pump}.
 In this case, for low temperatures compared to the charging energy,
 the only possibility for an electron to tunnel into the dot system
 is from the left lead to the left dot,
 leading to the transition from the state $|0,0\rangle$ to $|1,0\rangle$. It is
 described by the Lindblad operator $L_{\rm L}=|1,0\rangle\langle 0,0|$ and rate $\Gamma_{\rm L}$.
 Similarly, an electron can tunnel out into the right lead, creating a transition from $|0,1\rangle$
 to $|0,0\rangle$, which is described by $L_{\rm R}=|0,0\rangle\langle 0,1|$
 with rate $\Gamma_{\rm R}$.

 The pumping leads to a non-equilibrium state
 where the population in the state $|1,0\rangle$  with the higher energy is enhanced as compared
 to the state  $|0,1\rangle$ with lower energy. This effect persists when we go to the eigenbasis.
 For vanishing coupling to the radiation field the resulting population inversion becomes
 \begin{eqnarray}\label{eq:PI}
 \tau_0  &\equiv& \frac{\rho_{\rm st}^{\rm e} - \rho_{\rm st}^{\rm g}}
 {\rho_{\rm st}^{\rm e} + \rho_{\rm st}^{\rm g}}
  =\frac{\Gamma_{\rm R}\cos\theta-\Gamma_{\downarrow}}
 {\Gamma_{\rm R}
 \left[\displaystyle\frac{3+\cos(2\theta)}{4}\right]+\Gamma_{\downarrow}},
 \end{eqnarray}
where $\rho_{\rm st}$ denotes the steady-state density matrix and
$\rho_{\rm st}^{\rm e} = \sum_n \langle e,n|\rho_{\rm
st}|e,n\rangle$ the population of the excited state (tracing out the
degree of freedom of photon field), and similar for $\rho_{\rm
st}^{\rm g}$. Remarkably the expression for the ratio $\tau_0$,
obtained in eq. (\ref{eq:PI}) from a straightforward solution of the master equation
describing incoherent tunneling and relaxation,
does not depend on $\Gamma_{\rm L}$.

A positive population inversion, $\tau_0>0$, can lead to a lasing
state of the resonator. In the absence of relaxation, the population
inversion is maximized at $\theta=0$ where the excited and ground
states are simply the pure charge states $|1,0\rangle$ and
$|0,1\rangle$. However for this value, the coupling to the resonator $g$ vanishes.
 As we increase $\theta$ the coupling grows, 
 but the tunneling $t$ couples the pure charge states
more efficiently, and the population inversion decreases.
 At the degeneracy point ($\theta=\pi/2$) the population inversion vanishes.
Hence to achieve the lasing condition one should balance the two effects
and tune  $\theta$ to a value between
$0$ and $\pi/2$ and not too close to either limit.
Eq.~(\ref{eq:PI}) further shows how
relaxation processes, which transfer population from the excited
state to the ground state, reduce the population inversion.

Before proceeding we discuss the parameter regime considered in this
paper. To achieve the lasing state, the splitting between dot levels should be comparable to
the resonator frequency and the coupling $g$ strong enough to overcome the dissipation processes.
Considering the realistic bare coupling $g_0$ of the order of tens of MHz,
this requires both the bare energy difference $\epsilon$
and the interdot tunneling $t$ to be several $\mu{\rm eV}$ for a resonator with frequency in
the GHz regime.
In addition, the incoherent tunneling rate should be small, which we choose to be a few MHz.
For simplicity, we set $\Gamma_{\rm L} = \Gamma_{\rm R}=\Gamma$ throughout the paper.

\subsection{Ideal lasing conditions and sub-Poissonian statistics}

\begin{figure}[h]
\centering
\includegraphics[width=0.4\textwidth]{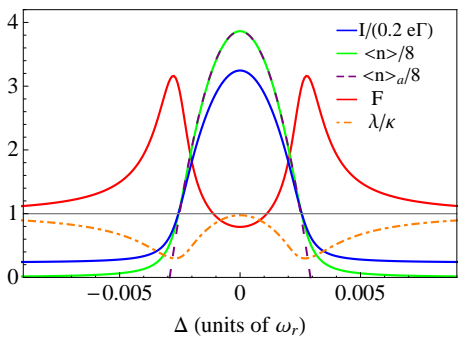}
\caption{(Color online) Average photon number $\langle n\rangle$,
compared with the analytical result $\langle n\rangle_{\rm a}$ in Eq. (\ref{eq_Approximate_result_for_n}),
Fano factor $F$, decay rate $\lambda$ and transport current $I$ as functions of the detuning
with $\Gamma = 10^{-3} \omega_{\rm r}$ and $t=0.3~\hbar\omega_{\rm r}$.
Throughout this paper, we choose the bare coupling strength $g_0 = 10^{-3}\omega_{\rm r}$.
 }\label{fig:Det}
\end{figure}

We first investigate the properties of the radiation field in the absence of
relaxation and decoherence processes of the two-level system,  $\Gamma_\downarrow=\Gamma_\varphi^*=0$.
We characterize the radiation field by
the average photon number $\langle n\rangle$, the Fano factor
$F$ as a measure of fluctuations, and the
decay rate $\lambda$ of the intensity correlation function (to be introduced later),
which are plotted in Fig.~\ref{fig:Det} as functions of the detuning $\Delta$.
The transport current, which will be discussed in the next section, is also shown in this figure for comparison.

The average photon number, which can be measured in experiments \cite{Astafiev},
has been investigated in detail in a single-qubit maser \cite{SQL1,SQL2,Didier}.
For large detuning, the quantum dots effectively do not interact with the resonator.
Thus the photon number in the resonator vanishes (at low $T$)
and the system is in the non-lasing regime.
Closer to the resonance, the system undergoes a lasing transition,
accompanied by a sharp increase in the photon number
which reaches a maximum at the resonance.
Approximately, the average photon number is given by \cite{Marthaler09},
\begin{eqnarray}\label{eq_Approximate_result_for_n}
 \langle n\rangle_{\rm a} \simeq \frac{\Gamma \cos\theta}{3\, \kappa}
 -\frac{\cos(2\theta)+7 }{96\, g^2}
 (4\Delta^2+\Gamma^2).
\end{eqnarray}
A comparison of this expression with numerical results, illustrated in
 Fig.~\ref{fig:Det}, demonstrates good agreement for large photon number.

The Fano factor $F \equiv (\langle n^2\rangle-\langle
n\rangle^2)/\langle n\rangle$, measures the deviation from a
Poissonian distribution, and hence the nature of the radiation field
\cite{fano}. In the non-lasing regime and in the absence of thermal
photons it can be approximated by $F\simeq\langle n\rangle +1$. When
the system approaches the lasing transition, the amplitude
fluctuations increase and the Fano factor grows. In the classical
lasing regime, the radiation field is in the coherent state and the
Fano factor $F=1$. As shown in Fig. \ref{fig:Det}, the Fano factor
can become smaller than 1, signalizing a sub-Poissonian distribution
of the radiation field. In this non-classical regime, the photon
number distribution is squeezed compared to the Poissonian
distribution \cite{Scully}.

Further information on the systems state is contained in
the intensity correlator $G(\tau) = \langle n(\tau) n(0)\rangle$,
which also can be measured in experiments.
It can be calculated using the quantum regression theorem \cite{Gardiner}
\begin{eqnarray}
 G(\tau) = {\rm Tr}[n\, e^{\mathcal L_{\rm tot}\tau} n \rho_{\rm st} ].
\end{eqnarray}
In the Markovian limit, the intensity correlator decays exponentially,
$G(\tau) - \langle n\rangle^2 \propto \exp(-\lambda \tau) $, with a
dominant decay rate $\lambda$. As shown in Fig.~\ref{fig:Det} the decay rate displays dips
when the system enters the transition regime
where the amplitude fluctuations are large and
the radiation field needs a long time
to relax back to the steady state.

\begin{figure}[h]
\centering
\includegraphics[width=0.35\textwidth]{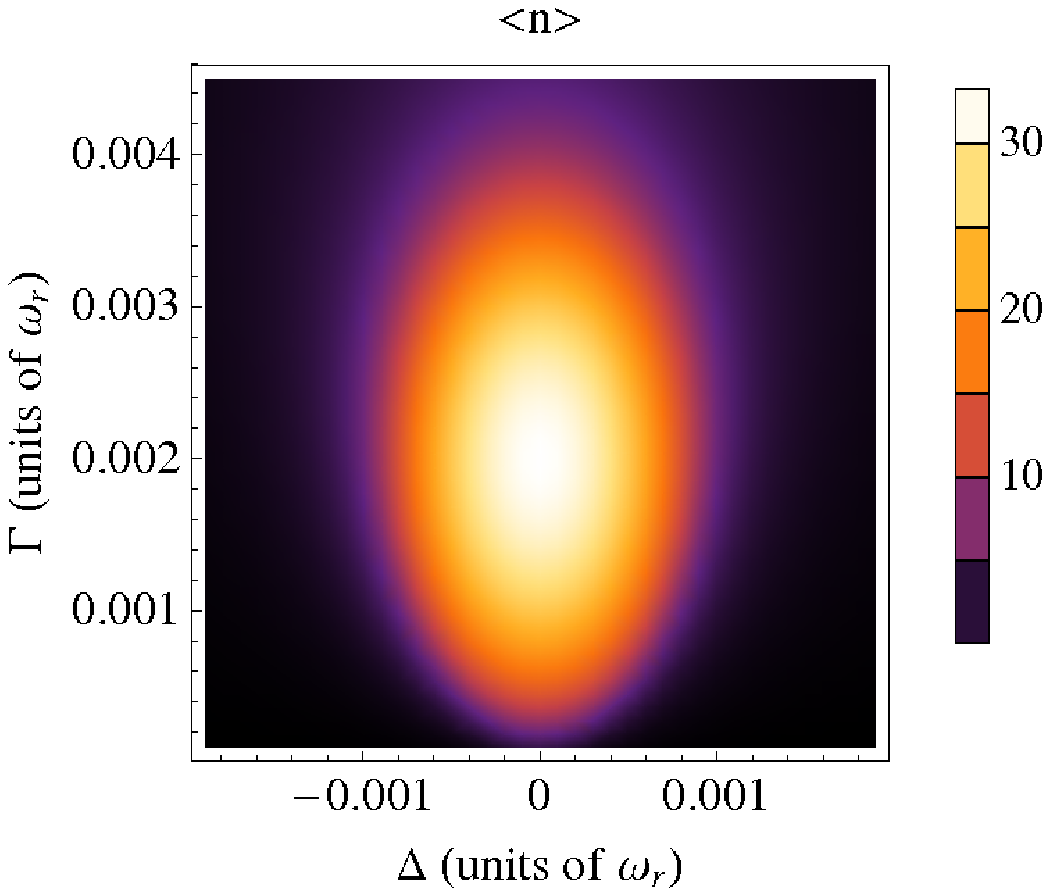}
\includegraphics[width=0.35\textwidth]{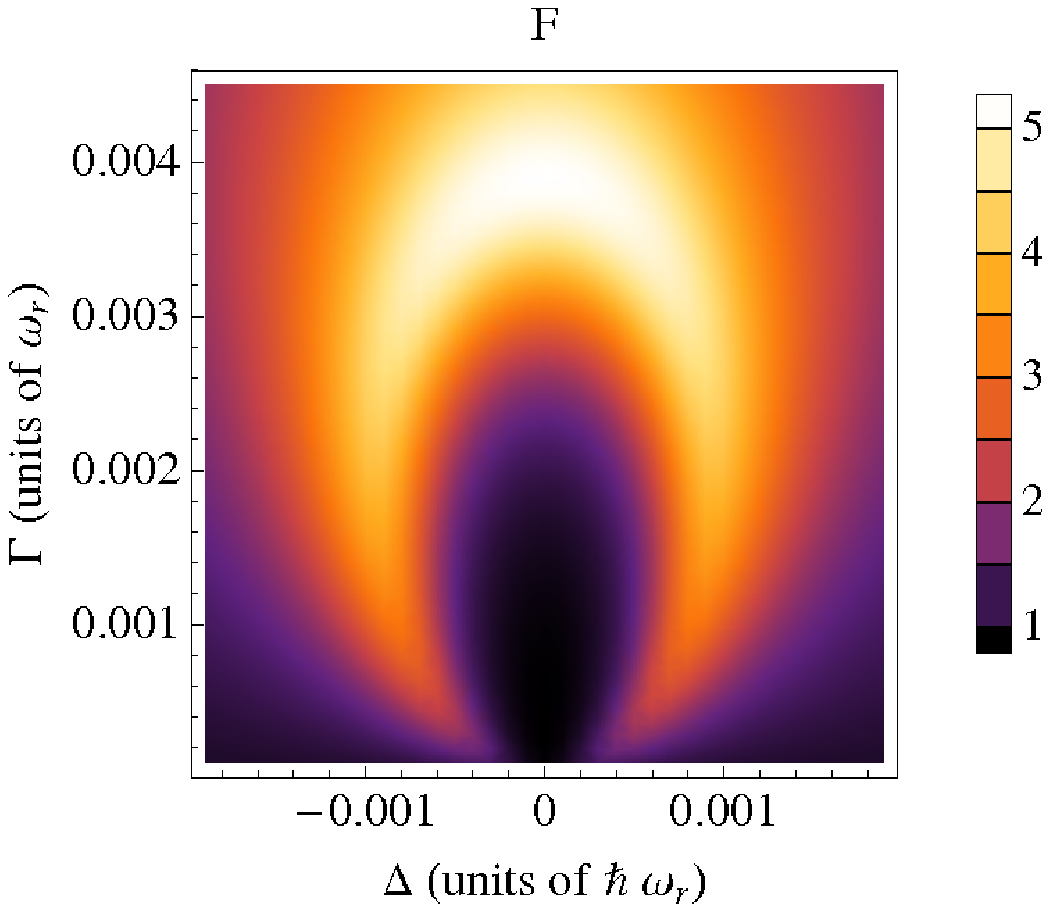}
\caption{(Color online) Average photon number and Fano factor versus
the bare energy difference $\epsilon$ and the incoherent tunneling
rate $\Gamma$ for interdot tunneling strength $t =
0.1~\hbar\omega_{\rm r}$ and vanishing relaxation and decoherence
rates in the dot system. }\label{fig:GammaL}
\end{figure}

Eq. (\ref{eq_Approximate_result_for_n}) also demonstrates the dependence
on the incoherent tunneling rate $\Gamma$, which can be tunned via gate voltages.
When $\Gamma$ is small compared to $g^2/\kappa$,
its pumping effect dominates and the photon number increases linearly with $\Gamma$.
When the incoherent tunneling becomes stronger, the decoherence caused by the incoherent tunneling
increases, which reduces the photon number.
This non-monotonous behavior is displayed in Fig. \ref{fig:GammaL}.

\subsection{Correlations between lasing and transport properties}

A current through the double dot requires that the tunneling cycle
$|0,0\rangle\rightarrow|1,0\rangle\rightarrow|0,1\rangle\rightarrow|0,0\rangle$
is completed. We can evaluate the current using the relation
\begin{eqnarray}\label{eq:curr}
 I = e\sum_{i,j}\Gamma_{i\rightarrow j}\,\langle i|\rho_{\rm st} |i\rangle,
\end{eqnarray}
where the index $i$ refers to the states $|g\rangle$, $|e\rangle$,
and $|0,0\rangle$, and $\Gamma_{i\rightarrow j}$ denotes the
transition rate from state $|i\rangle$ to $|j\rangle$.

\begin{figure}[t]
\centering
\includegraphics[width=0.38\textwidth]{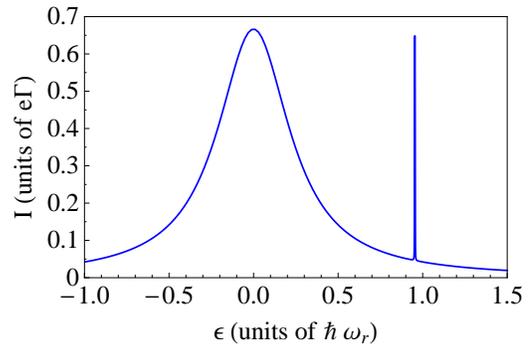}
\caption{(Color online) Current through the double-dot system 
as a function of the bare energy
difference $\epsilon$. 
The incoherent tunneling rate is $\Gamma =10^{-3}\omega_{\rm r}$, 
while the relaxation and decoherence are set to zero.
 We choose the interdot tunneling strength $t =
0.3~\hbar\omega_{\rm r}$, creating a nearly maximized population inversion
$\tau_0 \simeq 1$ as well as a strong enough coupling
strength $g\simeq 3\times10^{-4}\omega_{\rm r}$
to optimally satisfy the requirement (\ref{eq:lasing}).
The resonance with the oscillator occurs around
$\epsilon \simeq 0.95~\hbar\omega_{\rm r}$, leading to the 
sharp peak in the current. }\label{fig:curr}
\end{figure}

The resulting current, plotted as a function of the bare energy
difference $\epsilon$, is shown in Fig.~\ref{fig:curr}. We observe
two resonance peaks, since the coherent transition between the two
dots may be either elastic, or inelastic involving the creation of
an excitation in the resonator. The elastic transition leads to the
broad peak in the current around $\epsilon=0$ with width given by
the tunneling rate $t$ (here $t\gg
\hbar\,\Gamma$)~\cite{Vaart,Stoof}. 
A second peak appears at
$\epsilon = \sqrt{\hbar^2\omega_r^2-t^2}$, where the two-level system
is in resonance with the oscillator. Interestingly, this 
peak is -- for realistic values of the parameters
-- much narrower than the first one. It correlates with the lasing state, since the
excitation of a photon in the resonator is caused by the electron
tunneling between the two dots. The correlation between the lasing
state and the current is displayed in the results of Fig.
\ref{fig:Det}.

Both the lasing state and the current peak exist only in a narrow
 ``resonance window" $|\Delta|\le W/2$. From Eq. (\ref{eq_Approximate_result_for_n}) we find that
the condition $\langle n \rangle_{\rm a} \ge 0 $ yields
\begin{eqnarray}
 W = \Gamma \sqrt{\frac{32 \cos\theta\, g^2}{\left[\cos(2\theta)+7\right]\kappa\,\Gamma}-1},
\end{eqnarray}
which reduces to $W\approx 2 g\sqrt{\Gamma/\kappa-1}$ for small $\theta$.
For the estimate we could use that in the narrow resonance window
the mixing angle $\theta$ does not change much, and here, as well as for
the following discussions it is sufficient to fix $\theta$ to its value at the resonance.
For the considered parameter regime, the lasing window $W$ is around tens of MHz.

The height of the narrow current peak related to the lasing state can also be estimated.
For this purpose we adopt an adiabatic approximation assuming the dynamics of the resonator
to be much slower than that of the quantum dots \cite{Scully}.
For the considered parameter regime ($\kappa\ll \Gamma$ and small $\theta$),
the peak value of the current is then found to be
\begin{eqnarray}
 I(\Delta=0) \simeq e \Gamma \sum_{n=0}^{\infty} P(n)
 \left[\frac{2(n+1)}{3(n+1)+\Gamma^2/(4g^2)} \right].
\end{eqnarray}
Here $P(n)\simeq (\Gamma/\kappa) P(0)\Pi_{l=1}^{n} [3l+\Gamma^2/(4 g^2)]^{-1}$
denotes the probability of having $n$ photons in the resonator.
When the coupling to the resonator is strong compared to the incoherent tunneling $\Gamma$,
the peak current approaches $2e\Gamma/3$,
which for an incoherent tunneling rate of tens of MHz is of the order of pA.

The correlation between the lasing state and the transport current is remarkable in two ways.
On one hand, the current peak, which may be easier to measure than the photon number state of the
resonator, can be used as a probe of the lasing state. On the other hand, the rather
sharp resonance condition needed for the lasing makes the current peak narrow,
while at the same time the value of the current is reasonably high. This allows resolving
in an experiment even small details of the dot properties.
An example are the consequences of a difference in the Zeeman splittings
between the two dots, which is analyzed in a later part of the paper.

\subsection{Influence of relaxation and decoherence}

\begin{figure*}[t]
\hspace{5mm}
\includegraphics[width=0.38\textwidth]{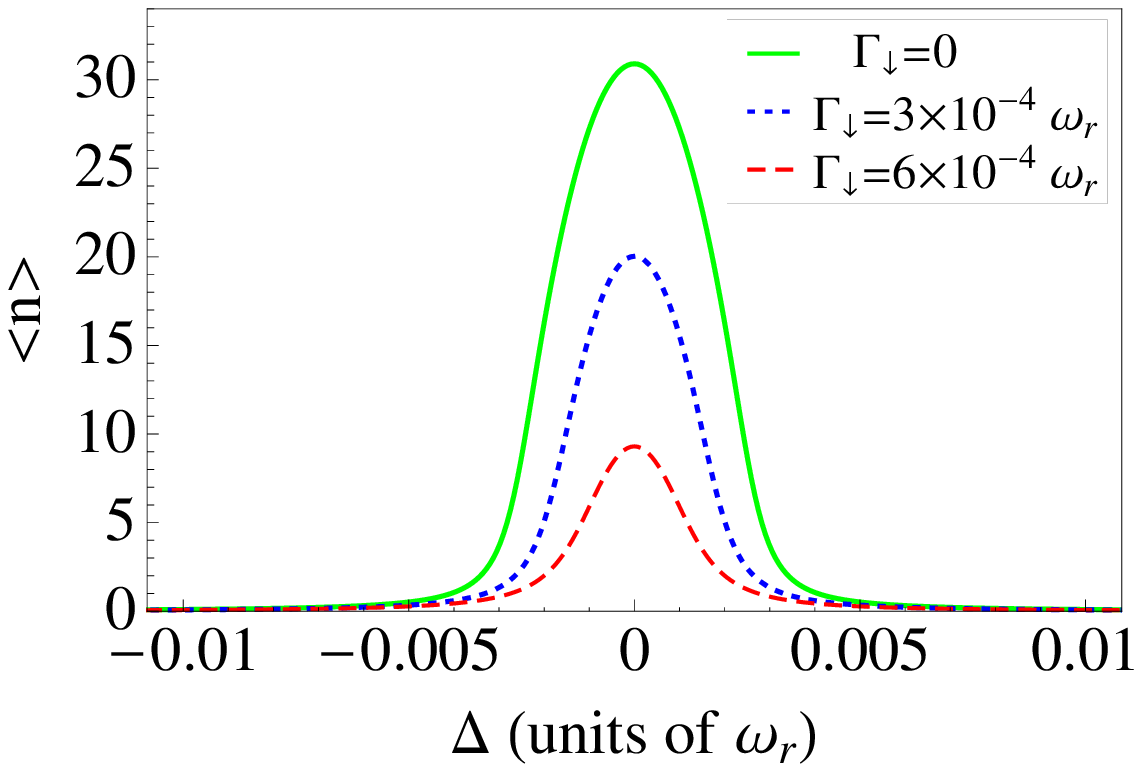}
\includegraphics[width=0.38\textwidth]{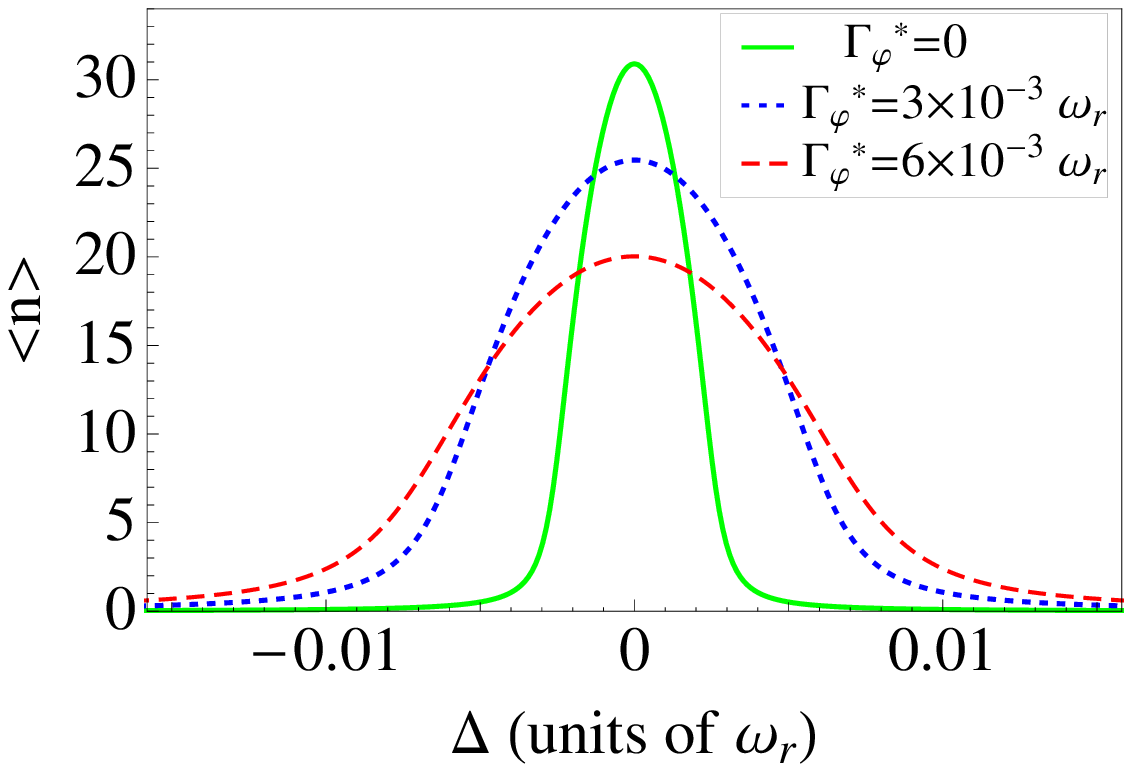} \hspace{5mm} \\[3mm]
\includegraphics[width=0.35\textwidth]{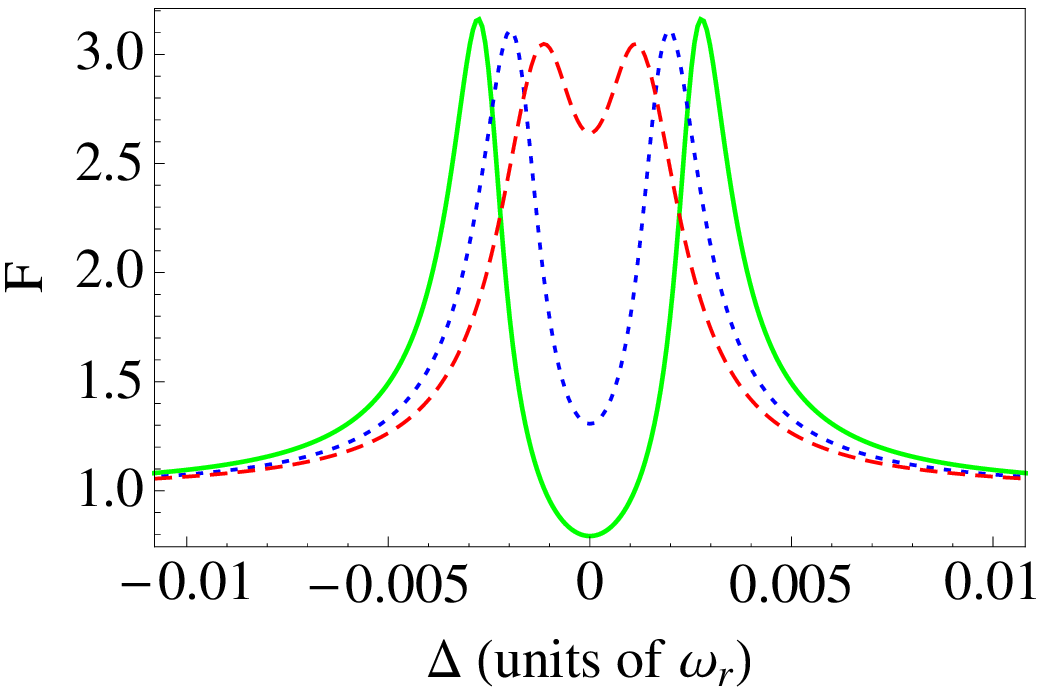} \hspace{3mm}
\includegraphics[width=0.35\textwidth]{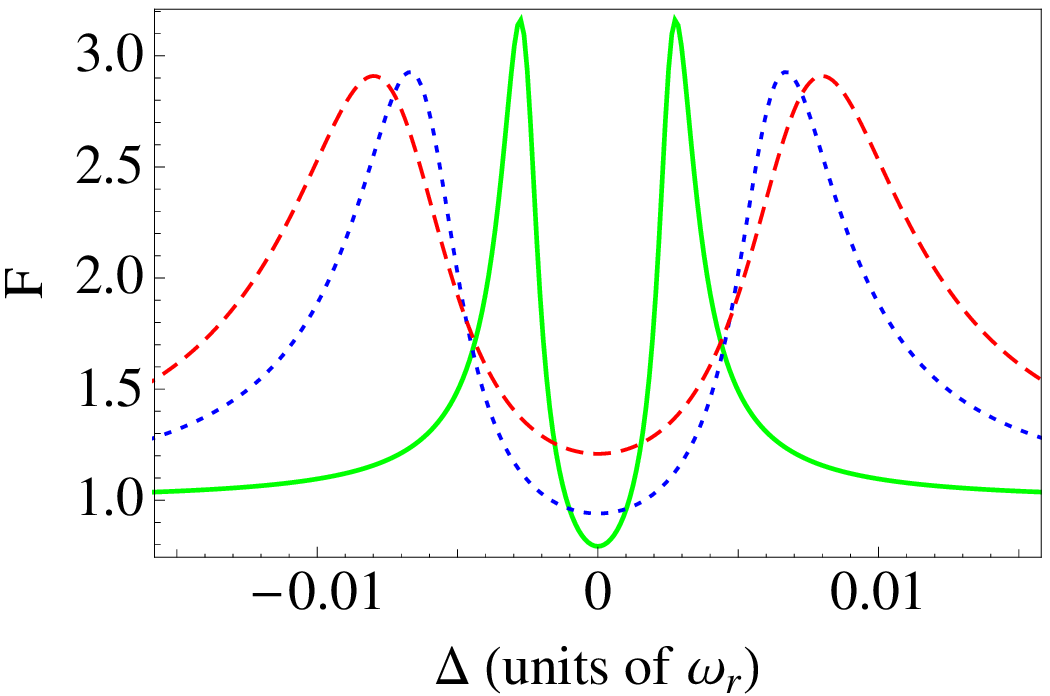} \\[3mm]
\includegraphics[width=0.35\textwidth]{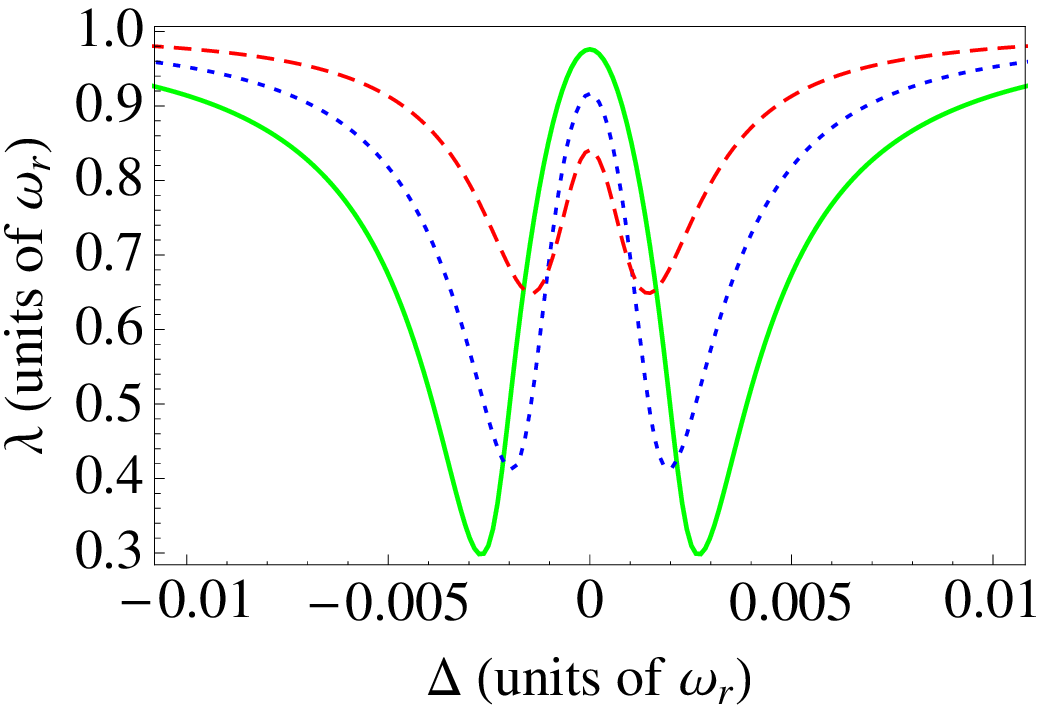} \hspace{3mm}
\includegraphics[width=0.35\textwidth]{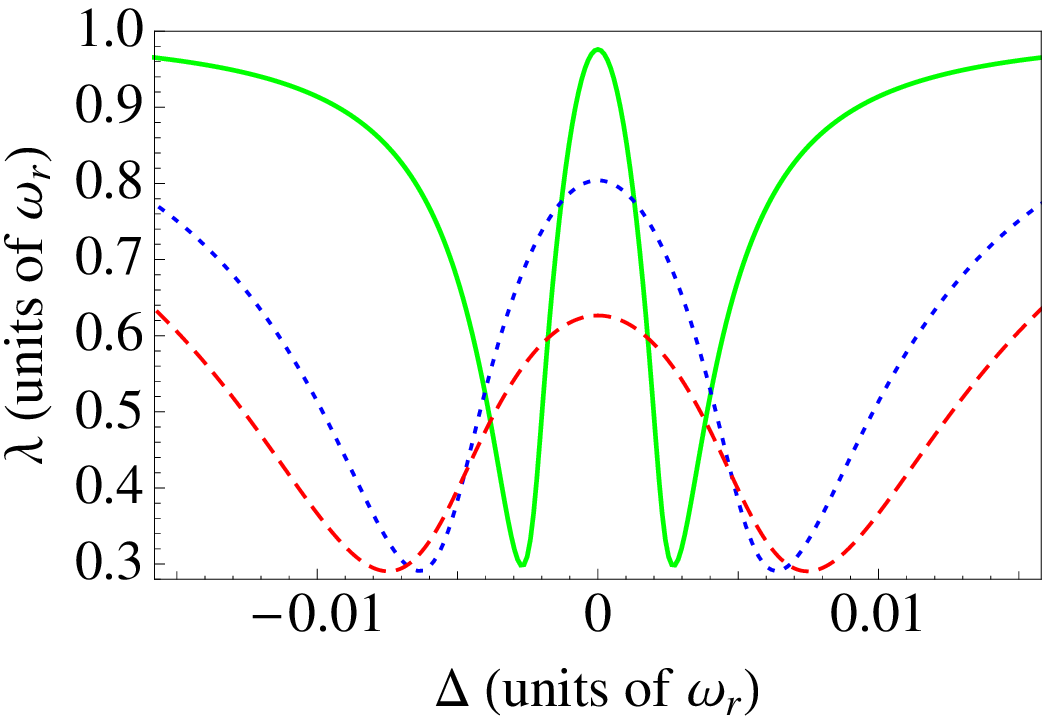} \\[3mm]
\includegraphics[width=0.35\textwidth]{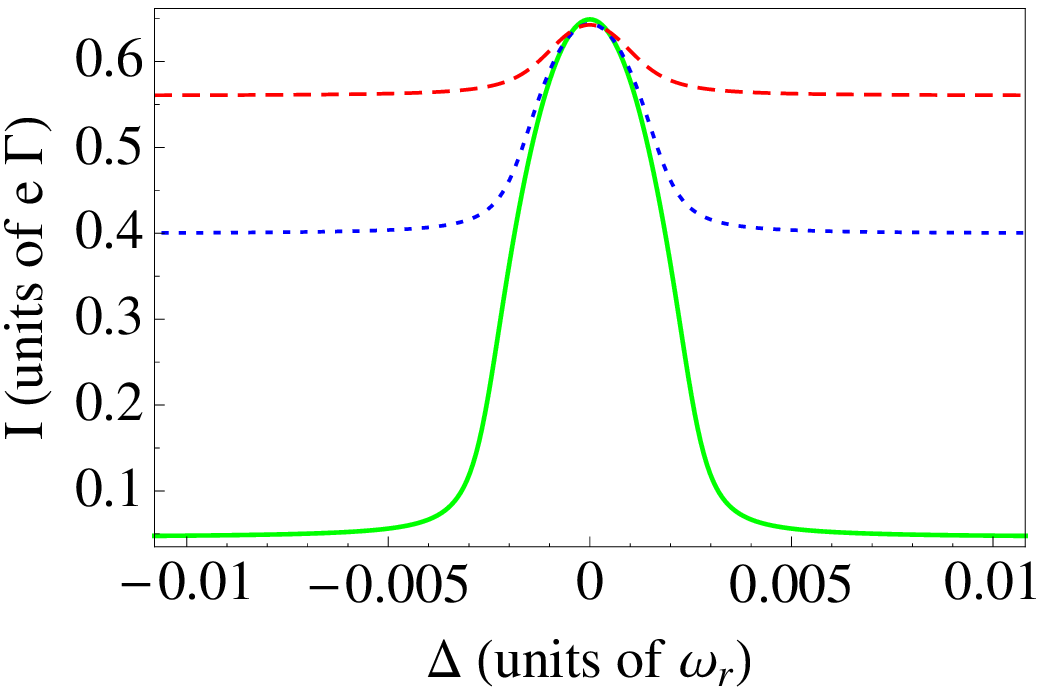} \hspace{3mm}
\includegraphics[width=0.35\textwidth]{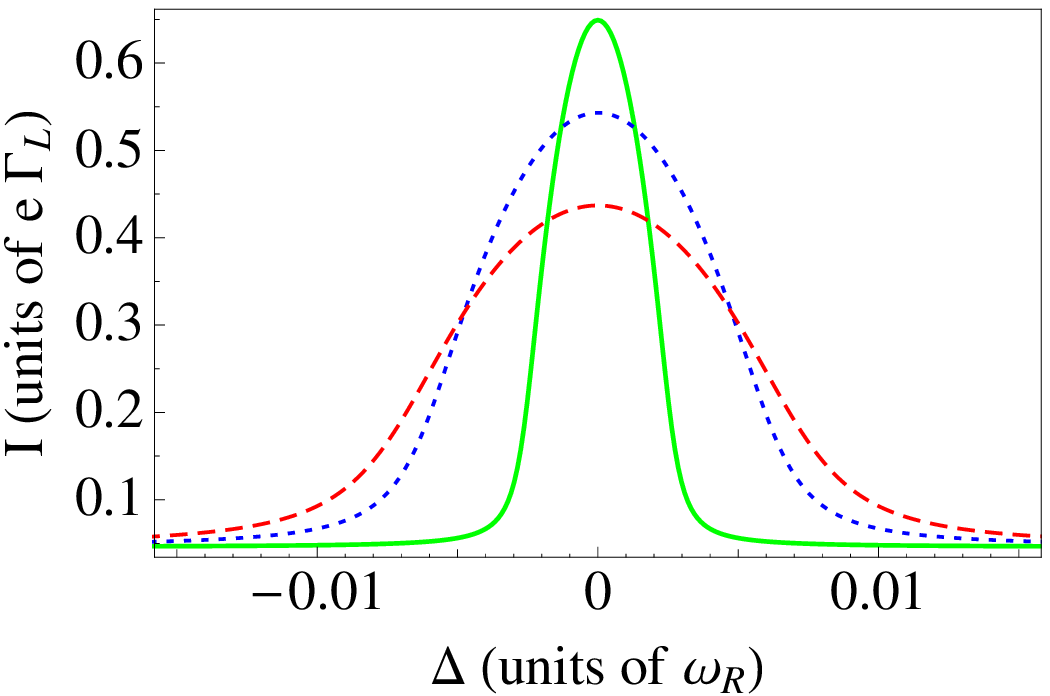}
\caption{(Color online)
Average photon number $\langle n \rangle$, Fano factor $F$, decay rate $\lambda$, and current $I$
through the double dot
as functions of the detuning for different relaxation rates but vanishing decoherence
in the left column, and for different decoherence rates but vanishing relaxation
in the right column. Other parameters are chosen to be the same in Fig. \ref{fig:curr}.
 }\label{fig:DecCharge}
\end{figure*}

In semiconductor quantum dot system dissipative effects causing relaxation
and decoherence are usually not negligible, and before proceeding we
have to analyze their influence.
Relaxation and decoherence of the two-level system modify the state
of the radiation field in different ways.
Figs.~\ref{fig:DecCharge} show the average photon number, the Fano factor,
the decay rate $\lambda$ and the transport current as functions of the detuning $\Delta$.
To illustrate the influence of relaxation and decoherence separately,
these quantities are plotted with different relaxation rates
without decoherence ($\Gamma_{\varphi}^* =0$) in the left column,
and with different decoherence rates but without relaxation ($\Gamma_{\downarrow} =0$) in the right column.

The results confirm that increasing the relaxation rate deteriorates the lasing state,
as expected already from Eq.~(\ref{eq:PI}), which shows that for stronger relaxation
the population inversion $\tau_0$ decreases.
The effects on the current are two-fold:
Relaxation reduces the efficiency of the coherent transition between the excited and ground states,
resulting in a decrease in both the current and photon number near the resonance.
On the other hand, relaxation also opens up an incoherent channel increasing the overall current.
As a result with growing relaxation rate, the current outside the resonance window becomes larger
and comparable to that inside the window.

Decoherence affects the lasing state in two ways. On one hand, it
has a destructive effect by decreasing the efficiency of energy
exchange between oscillator and the two-level system. On the other
hand, it effectively broadens the window in which the two-level
system can interact with the resonator. The competition between both
effects is illustrated in the right column of
Figs.~\ref{fig:DecCharge}. In the absence of decoherence, the
resonance window is narrow, but already a moderate decoherence rate
(e.g. $\Gamma_\varphi^* = 3\times10^{-3}~\omega_{\rm r}$), broadens
the resonance window while only modestly decreasing the average
photon number. For stronger decoherence, the width of the resonance
window is not significantly enhanced further, but destructive
effects shortening the lifetime dominate.


\section{Spin-split states}\label{Sec:Spin}

\subsection{Resolving Zeeman splitting difference via the lasing resonance}

In this section we extend the discussion to a situation where the
spin degeneracy is lifted by a magnetic field. When the Zeeman
splittings are the same for both quantum dots, the extensions are
straightforward. On the other hand, as we will see below, the properties of the system
are sensitive to even small differences in the Zeeman splittings in the two
dots $h_-=h_{\rm L} -h_{\rm R}$. The sharp resonance
condition for lasing allows resolving such small differences.
This may be useful in the context of quantum manipulations of electron spins in
quantum dots which serve as qubits.

In the presence of a magnetic field, which here is chosen to point
in $z$-direction, the spin degeneracy is lifted. As long as the
spin-orbit coupling is negligible (e.g., in GaAs quantum dots), the
$z$-component of spin is a good quantum number with eigenstates
denoted by $|\!\!\uparrow\rangle$ and $|\!\!\downarrow\rangle$.
Instead of two charge states, the relevant basis is now extended to
$|\!\!\uparrow,0\rangle$, $|0, \uparrow\rangle$,
$|\!\!\downarrow,0\rangle$, and $|0,\downarrow\rangle$, and the
Hamiltonian for the double dot system is given by
\begin{eqnarray}\label{eq:Hamiltonian_spin}
  H_{\rm dd} = \frac{1}{2} \left(
\begin{array}{cccc}
 \epsilon-h_{\rm L}  & t            &  0             &  0 \\
 t             & -\epsilon-h_{\rm R} &  0             &  0 \\
 0             & 0                  &  \epsilon+h_{\rm L} &  t \\
 0             & 0                  &  t             & -\epsilon+h_{\rm R} \\
\end{array} \right ),
\end{eqnarray}
where $h_{\rm L/R}$ represents the Zeeman splitting for the left/right dot.

\begin{figure}[t]
\centering
\includegraphics[width=0.38\textwidth]{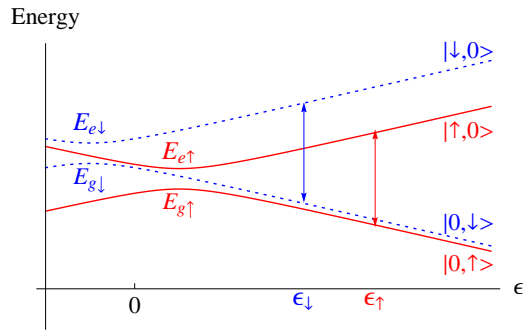}
\caption{(Color online) Eigenenergies for the spin-up channel,
$E_{\rm e\,\uparrow}$ and $E_{\rm g\,\uparrow}$ (red solid lines),
and spin-down channel, $E_{\rm e\,\downarrow}$ and $E_{\rm g\,\downarrow}$ (blue dotted lines),
in the presence of different Zeeman splittings in each dot.
The two spin channels become resonant with the radiation field
at separate values of the bare energy difference (denoted as $\epsilon_{\uparrow,\downarrow}$),
as indicated by the double-headed arrows.
For large $\epsilon\gg t$, the eigenstates of both channels
approach the basis states $|\!\!\downarrow,0\rangle$, $|0,\downarrow\rangle$, $|\!\!\uparrow,0\rangle$
and $|0, \uparrow\rangle$, as shown in the figure.
 }\label{fig:energyD}
\end{figure}

We assume the interdot tunneling to preserve the spin
and neglect spin flip-process induced by nuclear spins,
which  in GaAs quantum dots would be important only
at energy scales around $60$ neV~\cite{Petta05,Petta10}.
Hence, the Hamiltonian
(\ref{eq:Hamiltonian_spin}) separates into two decoupled spin channels with eigenstates
$|e_{\uparrow}\rangle$, $|g_{\uparrow}\rangle$, $|e_{\downarrow}\rangle$, and $|g_{\downarrow}\rangle$
and the corresponding energies shown in Fig. \ref{fig:energyD}.
Both spin channels can be coupled to the resonator via the charge dipole interaction,
in the same way as discussed above.
In the eigenbasis of both channels, the Hamiltonian for the coupled dot-resonator system
becomes
\begin{eqnarray}\label{eq:SpinH}
 H_{\rm sys} &=& \frac{\hbar\omega_{\uparrow} }{2} \sigma_z^{\uparrow}
 +\hbar g_{\uparrow} (a^\dag\sigma_-^{\uparrow}+a\sigma_+^{\uparrow})
 - \frac{h_+}{4} I_{\uparrow}
            \nonumber \\
 &+& \frac{\hbar\omega_{\downarrow} }{2} \sigma_z^{\downarrow}
 +\hbar g_{\downarrow} (a^\dag\sigma_-^{\downarrow}+a\sigma_+^{\downarrow})
 + \frac{h_+}{4} I_{\downarrow}
            \nonumber \\
 &+&\hbar\omega_{\rm r}\, a^\dag a,
\end{eqnarray}
with separate frequencies $\omega_{\uparrow/\downarrow} = \sqrt{(\epsilon\mp h_-/2)^2+t^2 }/\hbar$,
and Pauli matrices as well as identity matrices $I_{\uparrow/\downarrow}$ for the two spin channels.
E.g., we introduced $\sigma_z^{\uparrow} \equiv |e_{\uparrow}\rangle \langle e_{\uparrow}|
 -|g_{\uparrow} \rangle \langle g_{\uparrow}|$.
Here $h_\pm = h_{\rm L}\pm h_{\rm R}$, and  $h_-$ is assumed to take a positive value.

The difference in Zeeman splittings $h_-$ translates into different
frequencies for the two channels $\omega_{\uparrow/\downarrow}$.
The two spin channels become resonant with the radiation field
at different bare energy differences, which we denote as $\epsilon_{\uparrow/\downarrow}$
(shown in Fig.~\ref{fig:energyD}).
The distance between the two spin-resolved
resonances is simply the difference in Zeeman splittings,
$ \epsilon_\uparrow-\epsilon_\downarrow = h_-$.
For each spin-resolved
resonance, the photon number and current exhibit sharp peaks.
Therefore, if we can resolve the distances between the two peaks around the spin-resolved
resonances
we can extract the difference in Zeeman splittings.

\begin{figure}[t]
\hspace{2mm}
\includegraphics[width=0.4\textwidth]{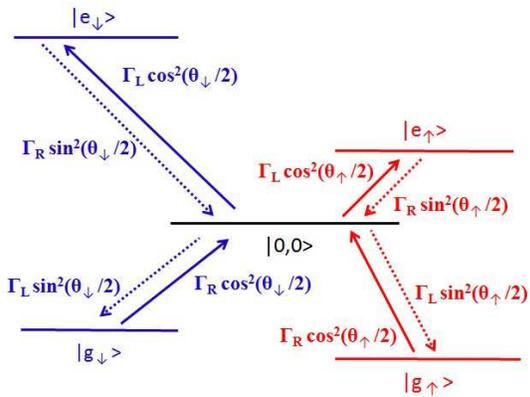}
\caption{(Color online) Pumping with spin-split states induced by a bias voltage across the double dots.
The eigenstates in both spin channels are connected to the common state $|0,0\rangle$
with transition rates depending on the mixing angles $\theta_{\uparrow/\downarrow}$.
For an effective pumping in both channels, the transitions from the ground state
$|g_{\uparrow/\downarrow}\rangle$ via $|0,0\rangle$
to the corresponding excited state $|e_{\uparrow/\downarrow}\rangle$,
denoted by solid arrows, are stronger than those flowing backward (dotted arrows).
The energy of state $|0,0\rangle$ is drawn at a position for a clear illustration.
}\label{fig:pumpspin}
\end{figure}

Before proceeding, we investigate the pumping with spin-split states,
which is illustrated in Fig. \ref{fig:pumpspin}.
For a bare energy difference $\epsilon$ in the vicinity of the spin-resolved resonances indicated in Fig. \ref{fig:energyD},
the transition rates from a ground state $|g_{\uparrow/\downarrow}\rangle$ to the common state $|0,0\rangle$
and then to its corresponding excited state $|e_{\uparrow/\downarrow}\rangle$ are stronger
than those of the backward processes. This asymmetry produces population inversions
in both spin channels.

\begin{figure}[t]
\centering
\includegraphics[width=0.38\textwidth]{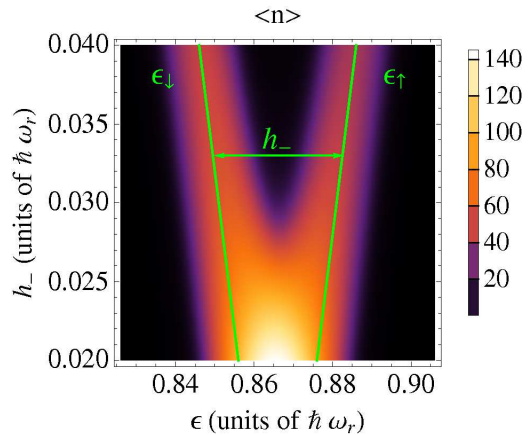} \\[3mm]
\includegraphics[width=0.38\textwidth]{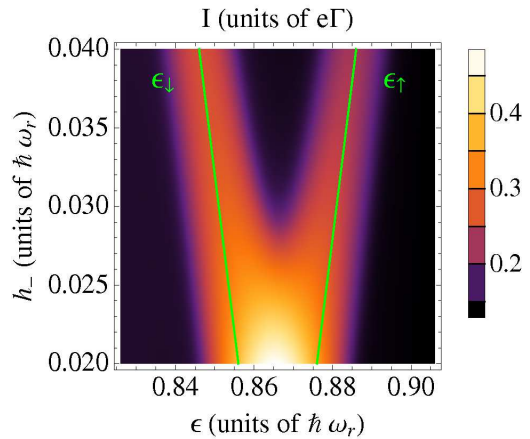}
\caption{(Color online) Average photon number and current in the
absence of both relaxation and decoherence. The two spin-resolved
resonances
$\epsilon_{\uparrow,\downarrow}$ are marked by the green solid
lines, which distance gives the difference in Zeeman splittings $h_-$.
In the discussion with spin-split states, we choose the interdot tunneling strength
$t=0.5~\hbar\omega_{\rm r}$ and incoherent tunneling rate
$\Gamma = 0.008~\omega_{\rm r}$. }\label{fig:SmallH}
\end{figure}

Fig. \ref{fig:SmallH} illustrates how the difference in Zeeman splittings
can be resolved in measurements of the photon number or current.
In the absence of both relaxation and decoherence,
two peaks can be resolved down to low values of $h_-/\hbar\omega_{\rm r}$.
For the transmission line resonator with GHz frequency,
this difference in Zeeman splittings corresponds to a magnetic field
around a few millitesla (for GaAs quantum dots),
which falls into the regime of a typical nuclear spin field \cite{Petta05}.
For smaller difference $h_-$, the two spin-resolved
resonances are close to each other,
where both channels couple to the radiation field inside one effective resonance window.
In this case, the photon number and current only show a single peak at the center
between the two spin-resolved
resonances
(denoted as $\epsilon_0$ for the purpose of later discussions).

\subsection{ Relaxation and population trapping }

Now we include the effects of relaxation and decoherence.
One would expect that relaxation deteriorates the ability to resolve
the states. However, at the spin-resolved
resonances the relaxation has a
constructive effect by increasing the population that can interact with
the radiation field. This effect even enhances the resolution of the
difference in the Zeeman splittings.

The reason for this constructive effect lies in the property that relaxation
releases trapped population during the pumping cycle. Here we
focus on one spin-resolved
resonance, e.g., $\epsilon_\downarrow$, to
illustrate the effect. As shown in Fig. \ref{fig:pumpspin}, the population in
the common state $|0,0\rangle$ can be pumped to the excited state of
either spin channel. For that pumped to the state
$|e_\downarrow\rangle$, it can be quickly transferred to the ground
state $|g_\downarrow\rangle$ by the resonant coupling to the
oscillator, and then back to the state $|0,0\rangle$. By contrast,
the population pumped to the state $|e_\uparrow\rangle$ is more
likely to be trapped there because of the less efficient coupling with
a finite detuning. As a result, when the system reaches the steady
state, most of the population is trapped in the state $|e_\uparrow\rangle$, and the
population that can interact effectively with the radiation field is
reduced.

\begin{figure}[t]
\includegraphics[width=0.38\textwidth]{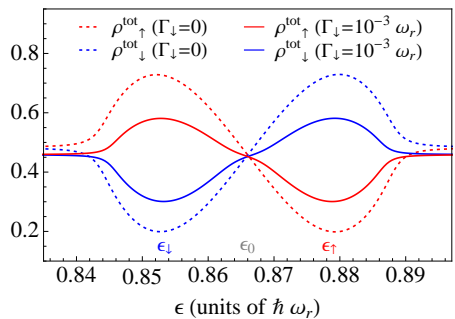}
\caption{(Color online) Total population in each spin channel
$\rho^{\rm tot}_{\uparrow/\downarrow}$ as a function of bare energy difference $\epsilon$.
Results obtained without relaxation are indicated by dotted lines
while those with a relaxation rate $\Gamma_\downarrow = 10^{-3}~\omega_{\rm r}$
by solid lines. Here we choose $h_- = 0.025~\hbar\omega_{\rm r}$.
}\label{fig:population}
\end{figure}

To further illustrate the effect of relaxation on the population trapped in
the off-resonant channel, we show in
Fig. \ref{fig:population} the total population
of each spin channel, namely, $\rho^{\rm tot}_{\uparrow/\downarrow} \equiv \rho_{e\uparrow/\downarrow}+\rho_{g\uparrow/\downarrow}$
with $\rho_{e\uparrow/\downarrow}$ and $\rho_{g\uparrow/\downarrow}$ being the population in the excited and ground states
in the corresponding channel.
At each spin-resolved
resonance, when the relaxation is absent, most of the population is trapped in the off-resonant channel.
This trapped population is partly released by relaxation to the resonant channel.
At the center $\epsilon_0$, the total population in each channel is
almost unchanged by the relaxation, since at this point, the two
channels are equivalent and no off-resonant channel exists to trap
the population.

\begin{figure}[t]
\hspace{5mm}
\includegraphics[width=0.4\textwidth]{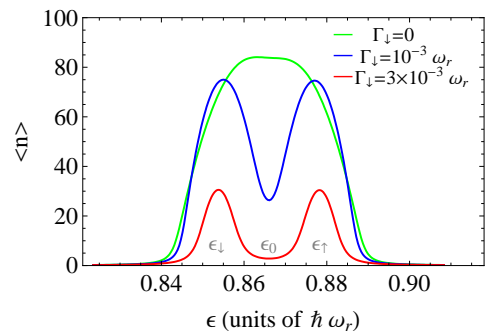}\\[3mm]
\includegraphics[width=0.36\textwidth]{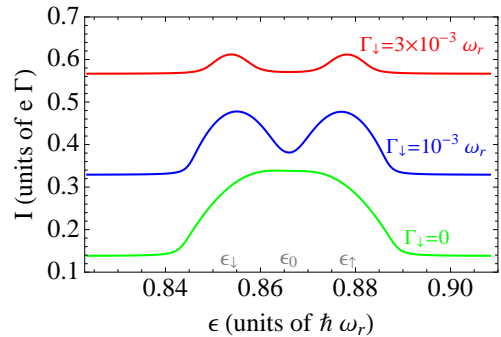}
\caption{(Color online) Average photon number and transport current
as functions of the bare energy difference $\epsilon$
with $h_- = 0.025~\hbar\omega_{\rm r}$ for different relaxation rates. }\label{fig:SpinRel}
\end{figure}

\begin{figure}[h]
\includegraphics[width=0.37\textwidth]{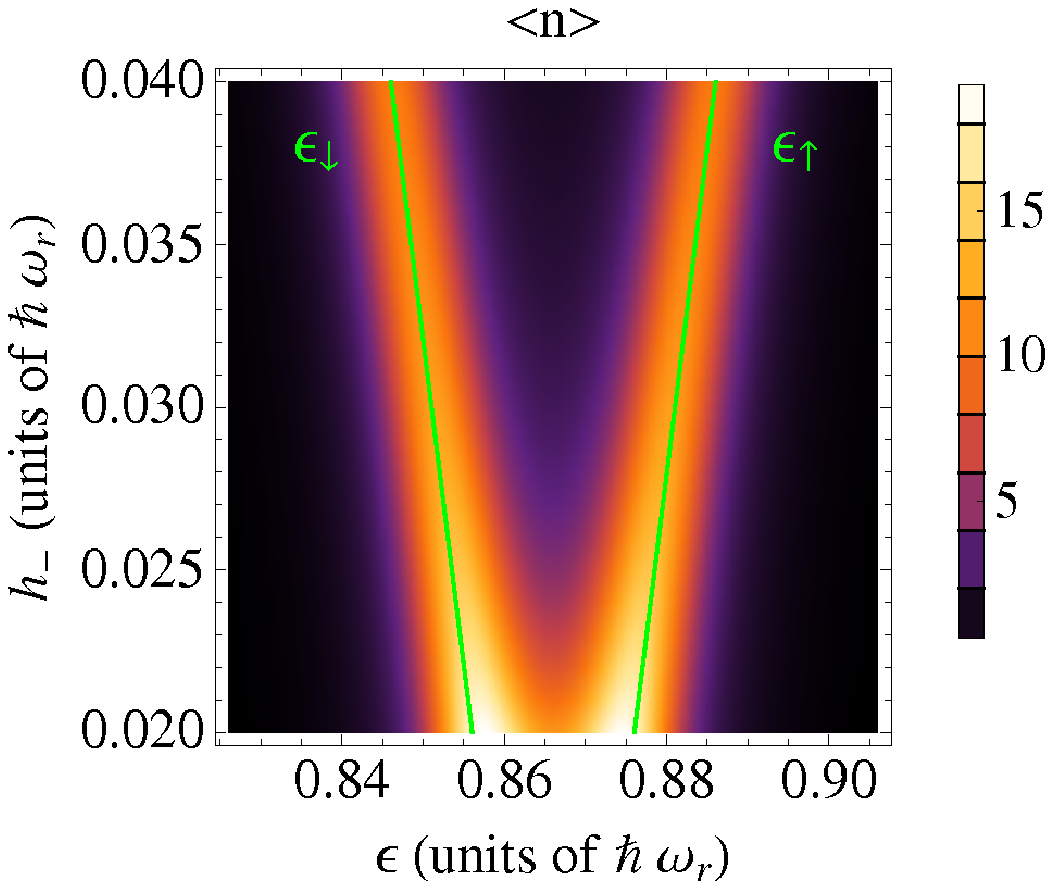}
\includegraphics[width=0.37\textwidth]{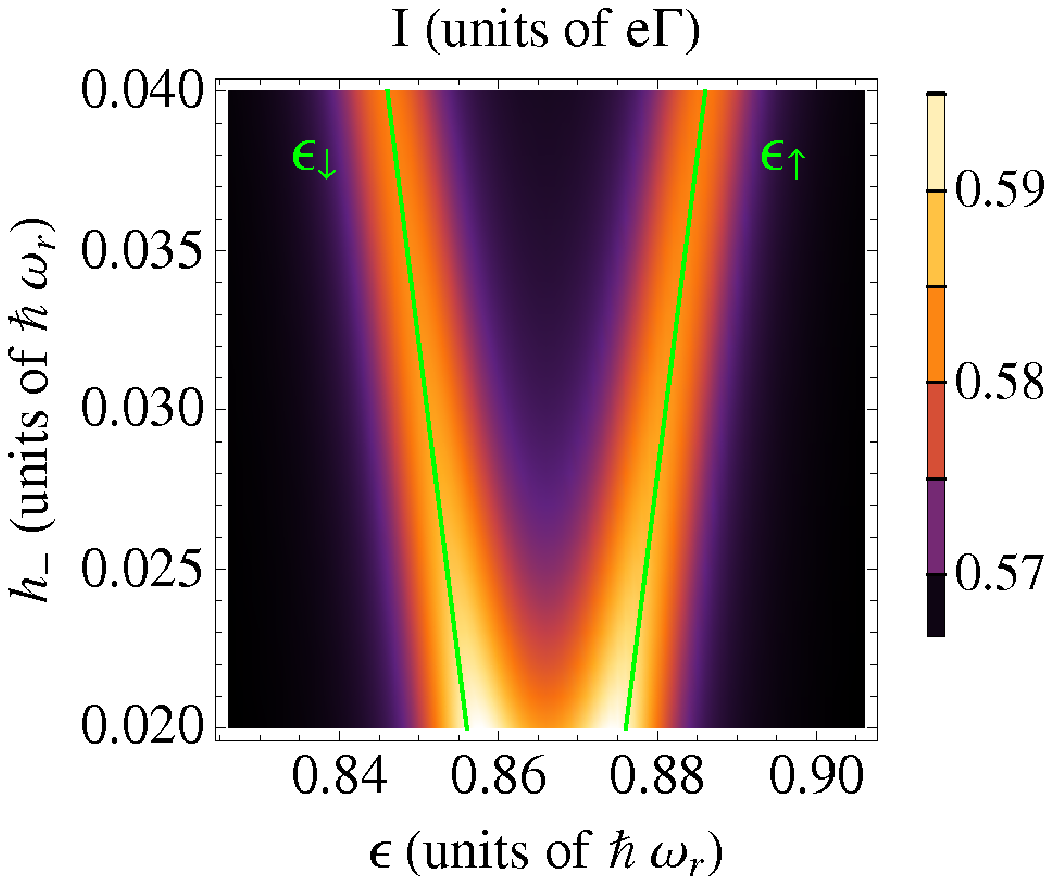}
\caption{(Color online) Average photon number and current in the
presence of both relaxation and decoherence with rates
$\Gamma_{\downarrow} = \Gamma_{\varphi}^* =
3\times10^{-3}~\omega_{\rm r}$. }\label{fig:NRelDec}
\end{figure}

For moderate strength relaxation, at each spin-resolved
resonance the effect of
releasing trapped population exceeds the destructive influences.
As shown in Figs. \ref{fig:SpinRel}, the photon number around the spin-resolved
resonances even slightly increase under the relaxation rate
$\Gamma_\downarrow = 10^{-3}~\omega_{\rm r}$. Around the center
$\epsilon_0$, the photon number drops significantly, where the
relaxation only has destructive influences. For strong relaxation,
the overall photon number drops. However, the two peaks in the
photon number are still resolvable under a relaxation rate
$\Gamma_\downarrow = 3\times10^{-3}~\omega_{\rm r}$, which
corresponds to a relaxation time around tens of nanoseconds with a
GHz resonator. For the current, since the relaxation also enhances the
current outside the resonance window, the peaks around the spin-resolved
resonances are less prominent. To further illustrate the effect and
show possible signatures in future experiments, we show in
Fig.~\ref{fig:NRelDec} the photon number and current in the presence of both
relaxation and decoherence.

\section{Summary}\label{Sec:Summary}

We have studied lasing and electron transport in a coupled quantum dot-resonator
system. To generate the population inversion required for the lasing,
we propose a pumping scheme induced by applying a voltage
across suitably biased double quantum dots. The pumping efficiency
increases as the system moves away from the degeneracy point. For an
optimal working point a balance between the pumping efficiency and
the coherent coupling to the resonator is required, since the
coupling becomes weaker away from the degeneracy point. The lasing
leads to a sharp current peak at the resonance with the resonator,
in addition to a broader peak arising from direct resonant interdot tunneling.
For realistic coupling strength to
the resonator, both peaks can be similar in height.

Relaxation processes shorten the life time of the two-level system and
reduce the population inversion and hence the lasing effect.
Decoherence, which also shortens the life time,
broadens the effective resonance window between the quantum dots and resonator.
For the transport current, the relaxation opens up an extra incoherent channel
which increases the current outside the resonance window with the resonator.

The sharp resonance condition allows for resolving small
differences in the dot properties. This opens perspectives for applications of the
setup and operation principle for high resolution measurements.
As an example we studied the consequences of different Zeeman splittings between the two dots.
Two separate peaks arise in both the photon number and current with
distance given by the Zeeman splitting difference.
Due to the narrow resonance condition for lasing
the two peaks can be resolved for realistic parameters.
Relaxation processes can
even enhance the resolution by releasing population trapped in the off-resonant spin channel.
For relaxation and decoherence with rates around a
hundred of MHz, a resolution down to a few millitesla can be achieved.

\section*{Acknowledgements}
We acknowledge helpful discussion with S. Andr\'e, A. Romito, and J. Weis,
as well as the financial support from the Baden-W\"{u}rttemberg Stiftung
via the Kompetenznetz Funktionelle Nanostrukturen.

\end{document}